# Visualizing flat-band spatial renormalization in rhombohedral graphene superlattices

Peng-Cheng Pan[1, 2,†], Shihao Zhang[1,†], Yang Zhang[1, 2], Ji Huang[1, 2], Ling-Hui Tong[1, 2], Chen-Chen Xu[1, 2], Yuan Tian[1], Li Zhang[1], Lijie Zhang[1], Yuanyuan Hu[3], Wen-Xiao Wang[4,*], Zhihui Qin[1,*], and Long-Jing Yin[1, 2,*]

[1] *Key Laboratory for Micro/Nano Optoelectronic Devices of Ministry of Education & Hunan Provincial Key Laboratory of Low-Dimensional Structural Physics and Devices, School of Physics and Electronics, Hunan University, Changsha 410082, China*

[2] *Yuelushan Center for Industrial Innovation, Changsha 410082, China*

[3] *College of Semiconductors (College of Integrated Circuits), Hunan University, Changsha 410082, China*

[4] *College of Physics and Hebei Advanced Thin Films Laboratory, Hebei Normal University, Shijiazhuang 050024, China*

[†]These authors contributed equally to this work

*Corresponding author: wangwx@hebtu.edu.cn; zhqin@hnu.edu.cn; yinlj@hnu.edu.cn

**Rhombohedral graphene/hBN moiré superlattices exhibit flat-band-driven emergent phases, including superconductivity and the fractional quantum anomalous Hall effect (FQAHE), yet the microscopic role of the moiré potential remains unclear. Here, using scanning tunneling microscopy, we visualize moiré-modulated spatial renormalization of flat bands in rhombohedral pentalayer and tetralayer graphene/hBN superlattices. We observe spatially hierarchical filling, manifested as periodic energy shifts of the flat bands at the moiré scale, leading to spatial reshaping of correlated states in the interacting regime. Remarkably, this modulation vanishes below a ~10 nm moiré period—the same threshold below which the FQAHE is absent. Theoretical modeling attributes this mechanism to atomic-corrugation-induced charge redistribution. Our work provides real-space visualization of moiré-engineered flat-band reconstruction, resolving a key link between moiré periodic potential and emergent topological order.**

Rhombohedral graphene (RG) has emerged as an ideal model system for investigating correlated and topological physics, owing to its intrinsic low-energy flat electronic bands that feature tunable nontrivial topology (*1-9*). In this system, both pristine RG and its moiré superlattice formed by alignment with hexagonal boron nitride (hBN) have been found to host a wealth of analogous yet intricate emergent quantum phenomena driven by correlation and topology (*10-18*), including unconventional superconductivity (*19-23*), magnetism (*24-28*), and the integer and fractional quantum anomalous Hall effects (IQAHE and FQAHE) (*29-34*). Among these, the FQAHE stands out: it is a newly discovered Hall effect (*35-38*) and is regarded as a promising platform for exploring non-Abelian anyons and topological quantum computation. Interestingly, the FQAHE in RG has been observed exclusively in hBN-aligned moiré systems—first in pentalayer, then in tetralayer and hexalayer graphene (*29, 33, 34*)—all with a moiré period $\lambda_\mathrm{m} \gtrsim 10$ nm, strongly suggesting a critical role of the moiré potential. Yet contrary to expectations based on conventional moiré engineering, transport experiments reveal that the FQAHE emerges preferentially on the moiré-distant side of the RG/hBN device—where electrons are polarized away from the moiré interface—rather than on the moiré-nearby side. This puzzling observation raises a fundamental question: what is the actual role of the moiré potential in forming the FQAHE? Resolving this question is essential for understanding not only the FQAHE but also moiré-driven emergent states in graphene more broadly. Although several theoretical proposals have attempted to explain this discrepancy, they remain contradictory (*39-43*). What has been missing is a real-space, microscopic investigation of how the moiré potential reshapes the flat bands in RG.

Here, using gate-tunable scanning tunneling microscopy and spectroscopy (STM/STS), we directly image the spatial moiré modulation of flat bands on the moiré-distant side of rhombohedral pentalayer and tetralayer graphene (R5G and R4G) aligned with hBN. We uncover a spatially renormalized flat-band landscape that varies periodically with the moiré superlattice. At full and empty flat-band fillings, both the flat bands and remote bands exhibit periodic energy shifts at the moiré scale. At partial fillings, we observe a striking periodic reshaping of correlation-induced flat-band splitting. Remarkably, this entire moiré-modulated pattern disappears when the moiré period falls below ~10 nm—exactly the threshold below which the FQAHE is absent. Combined with spectroscopic mapping and theoretical modeling, our work reveals a spatially hierarchical filling rule driven by atomic-corrugation-induced intralayer charge redistribution in the RG moiré system.

**Single-particle spectroscopy of RG/hBN moiré superlattices**

Our RG/hBN superlattice devices were fabricated on $SiO_2$/Si substrates with a back gate connected to the silicon (see Fig. 1A for the experimental setup and figs. S1 and S2 for more details of sample fabrications). The STM/STS measurements were performed at $T \approx 5$ K unless otherwise stated. A moiré superlattice is formed due to the lattice mismatch of ~1.7% between graphene and hBN (*44*) and has a maximum period of ~15 nm when they are stacked in alignment (i.e., twist angle $\theta = 0°$). Figure 1, C and D, shows the STM topography of R4G/hBN (device A) and R5G/hBN (device B) superlattice samples, respectively. The moiré pattern is clearly visualized in the images and exhibits a surface-clean, homogeneous feature that can extend over hundreds of nanometers. Similar topography was observed in all samples, indicating the high quality of our devices. There are three high-symmetry stacking sites $C_{BN}$, $C_B$, and $C_N$ within the moiré unit cell of RG/hBN superlattice (as labeled on the STM images in Fig. 1, C and D). The $C_{BN}$ site represents the configuration where carbon atoms of the bottom graphene layer sit directly atop both boron and nitrogen atoms of the hBN, while $C_B$ and $C_N$ sites indicate configurations where a single carbon atom lies directly above a boron atom or a nitrogen atom, respectively (see Fig. 1B for illustration of the stacking configurations). These local stacking configurations possess distinct stacking energies, which in turn lead to different local interlayer repulsion forces (*44-46*). As a result, they display varying topographic heights in the STM image. In particular, the $C_B$ configuration has the lowest stacking energy and, thus, typically demonstrates an extended region within the moiré unit cell at small twist angles due to lattice relaxations (*47, 48*).

Figure 1, E and F, shows the representative *dI/dV* spectra—reflecting the local density of states (LDOS) of electrons at the sample surface—measured on the R4G/hBN and R5G/hBN superlattices, respectively. In both systems, sharp DOS peaks arising from the RG flat bands are clearly resolved near the Fermi level (zero bias), accompanied by remote band features at approximately ±0.3 eV. These spectroscopic signatures are consistent with those observed in pristine R4G and R5G (*12, 15, 16*), indicating that the intrinsic flat-band physics of RG remains robust in the presence of the hBN moiré potential. However, the moiré potential also introduces new effects. Unlike in pristine RG, where the valence flat band (VFB) and conduction flat band (CFB) are degenerate at zero displacement field ($D = 0$), the moiré superlattice lifts this degeneracy, leading to a finite splitting between VFB and CFB even in the absence of an applied interlayer bias. As shown in Fig. 1E, at $D = 0$ we observe two flat-band peaks separated by ~18 meV with comparable spectral weight. When a finite displacement field is applied—either via back-gate voltage or intrinsic auto-doping from the device (*49, 50*)—the VFB and CFB further separate and become polarized toward the top and bottom layers (see figs. S3 and S4 for detailed determination of $D$). This layer polarization manifests in the $dI/dV$ spectra as two highly asymmetric peaks (Fig. 1F), because STS primarily probes the electronic wavefunctions of the top-layer-resided (moiré-distant) flat band (*16, 51, 52*). These single-particle behaviors are well captured by our spectroscopic measurements, which show excellent agreement with theoretical calculations (Fig. 1, G and H). The above spectroscopic results establish a foundation for understanding the more complex moiré and correlated effects in RG/hBN superlattices, to which we turn below.

**Moiré-period-governed spatial band renormalization**

We now investigate the spatial modulation of flat bands by the moiré effect, beginning with the non-interacting regime—that is, at non-partial fillings of the flat bands. Figure 2, B and C, presents spatially resolved $dI/dV$ spectra measured across several moiré unit cells of a $\theta = 0.23°$ R5G/hBN superlattice (Fig. 2A; $D \neq 0$) at full filling of moiré-distant flat band. Remarkably, the flat-band position exhibits a pronounced site-dependent energy shift. Specifically, it shifts downward at both the $C_{BN}$ and $C_N$ sites, revealing a spatial oscillation that faithfully follows the moiré periodicity (Fig. 2, B to D). Moreover, nearly identical behavior was also observed when the flat band is fully unoccupied, as shown in Fig. 2, F to I, of a $\theta = 0.66°$ R5G/hBN sample. In this case, the flat band again displays a clear moiré-site-dependent shift, confirming that the effect is robust regardless of filling polarity.

Collectively, the above observations establish the existence of moiré-induced spatial renormalization of flat bands in RG/hBN superlattices. The renormalization strength, defined as the maximum energy shift between the $C_{BN}$ and $C_B$ sites, is approximately 15 meV for the samples shown in Fig. 2, A and F—comparable in magnitude to the calculated moiré potential of graphene/hBN system (*44, 48, 53*). Notably, we also observed that the remote bands shift in the same direction and by a similar magnitude as the flat bands (Fig. 2, E and J). This parallel behavior suggests that the observed energy shifts are not specific to the flat bands themselves but rather reflect a moiré-modulated variation of the local chemical potential. In other words, the moiré effect induces spatial inhomogeneity in local doping, which globally shifts all electronic bands in concert.

Interestingly, this moiré-scale band renormalization is absent in RG/hBN devices with smaller moiré periods. Figure 2, K to N, shows site-dependent *dI/dV* spectra recorded in a $\theta = 1.29°$ R4G/hBN superlattice ($\lambda_m \approx 8.8$ nm). In stark contrast to the large-period case, the flat-band position remains highly homogeneous across different moiré sites, indicating negligible spatial modulation. To systematically map out the period-dependent behavior, we measured a series of R4G and R5G moiré superlattice samples with periods ranging from 7.7 nm to 14.4 nm (see figs. S5 to S7 for details of the data). As summarized in Fig. 2O, the flat-band offset emerges only when the moiré period $\lambda_m$ exceeds approximately 10 nm, corresponding to a twist angle $\theta \lesssim 1°$. This explicitly demonstrates a distinct moiré-periodicity–governed band renormalization.

Remarkably, the observed period threshold coincides nearly exactly with the condition under which the FQAHE has been reported in transport studies of RG/hBN superlattices (*29, 33, 34*), and both occur on the moiré-distant side. This quantitative agreement provides a crucial microscopic perspective for assessing the actual role of the moiré effect in forming the FQAHE—a central theme we address below.

**Spatial reshaping of correlated states**

We next turn to the interaction regime, examining how the moiré effect influences electronic correlations on the moiré-distant side. Figure 3 shows spatially resolved *dI/dV* data measured on RG/hBN samples with $\lambda_m > 10$ nm ($\theta = 0.38°$) and $\lambda_m < 10$ nm ($\theta = 1.29°$) when the flat bands are partially filled. In both cases, the flat-band peak exhibits a clear correlation-induced splitting near the Fermi energy (see fig. S3 and S8 for details). However, the spatial evolution of this splitting differs dramatically between the two regimes. For the $\lambda_m < 10$ nm sample (Fig. 3, G to J; see fig. S9 for more data), the flat-band splitting remains nearly uniform across different moiré sites and unit cells, with a constant gap of ~15 meV. In contrast, for the $\lambda_m > 10$ nm sample (Fig. 3, B to D), the splitting gap exhibits a periodic oscillation that faithfully follows the moiré pattern. This spatial modulation is more vividly visualized in the gap-size map (Fig. 3E), where a clear moiré-scale pattern of the splitting gap emerges. This moiré-modulated reshaping of the flat-band splitting was consistently observed across multiple RG/hBN superlattice samples with $\lambda_m > 10$ nm (figs. S10 and S11), indicating that the correlated states themselves inherit the spatial texture imposed by the moiré effect. Notably, the parameter space in Fig. 3, A to E ($D \approx 0.7$ V/nm) closely matches that of transport experiments reporting the QAHE ($D \approx 0.7$-$1.0$ V/nm) (*29, 33*), indicating that the moiré-modulated spatial reshaping of correlated states can persist at large-*D* regime.

**Hierarchical filling visualized in real space**

Collectively, the above results demonstrate a moiré-periodicity-governed spatial renormalization of flat bands in RG/hBN. In small-period superlattices ($\lambda_m \lesssim 10$ nm, $\theta \gtrsim 1°$), the spatial band renormalization is negligible. In large-period systems ($\lambda_m \gtrsim 10$ nm, $\theta \lesssim 1°$), however, the moiré superlattice substantially modulates the local doping level of the moiré-distant flat band, giving rise to a spatially hierarchical filling sequence at the moiré scale. Because correlation effects are highly sensitive to local filling, this spatially varying doping can reshape the correlated states in real space when interactions are turned on.

This hierarchical filling mechanism is further corroborated by real-space spectroscopic mapping of electronic states. Figure 4 shows a series of STS maps acquired around the moiré-distant flat band for RG/hBN devices with different periods. For $\lambda_m > 10$ nm samples, both at full filling (Fig. 4, A to C) and empty filling (Fig. 4, D to F) of the flat band, the surface electronic states exhibit clear moiré-scale spatial modulation. On the left side of the flat-band peak, the electronic states are predominantly localized at the $C_{BN}$ and $C_N$ sites, forming a hexagonal honeycomb-like moiré pattern that imitates the topological morphology. As the energy sweeps to the right side of the flat-band peak, the electronic states gradually transfer to the $C_B$ site and eventually become localized there, producing a complementary moiré pattern relative to the preceding maps. This systematic energy-dependent spatial migration of flat-band LDOS was observed across various large-period RG/hBN superlattices (figs. S12 to S14). In sharp contrast, for $\lambda_m < 10$ nm ($\theta \gtrsim 1°$) samples, this phenomenon is virtually absent. As shown in Fig. 4, G to I, for a $\theta = 1.41°$ R4G/hBN device, the electronic states across different energies—including at the flat band—remain relatively uniform in real space (see fig. S15 for more data), indicating negligible moiré modulation. These real-space mappings directly visualize a moiré-periodicity-governed, spatially hierarchical filling phenomenon at the moiré-distant surface of RG/hBN superlattices.

**Theoretical mechanism of spatially hierarchical filling**

To elucidate the underlying origin of the moiré-modulated spatial filling hierarchy, we performed twist-angle-dependent calculations. Our modeling points to moiré corrugation as the governing factor. Figure 5 shows the out-of-plane and in-plane atomic corrugations of the topmost graphene layer in the R5G/hBN system with different twist angles. This moiré corrugation can give rise to deformation potential, tuned interlayer hopping, and electrostatic energies induced by the hBN substrate. For the topmost graphene layer—the moiré-distant surface—the interlayer distance between adjacent graphene layers remains almost unchanged, as these layers are far from the hBN substrate. Therefore, the effect of tuned interlayer hopping can be ignored when analyzing the real-space-dependent on-site potential. Electrostatic energies induced by the hBN substrate depend on the relative stacking between RG and hBN in real space, and this effect exists in all RG/hBN samples regardless of twist angle. Thus, the contribution of hBN-substrate-induced electrostatic energies can be ruled out when discussing the twist-angle-modulated behavior. This leaves atomic corrugation as the primary origin of the site-dependent on-site energy variation of the flat bands.

Atomic corrugation generates both deformation potential and pseudo-magnetic field. The pseudo-magnetic field induces a k-shift of the Dirac points but does not affect the on-site potential tuning of the flat bands. The deformation potential, by contrast, originates from non-uniform spontaneous atomic strain. Our results show that at small twist angles ($\theta < 1°$; Fig. 5, A to C), both out-of-plane and in-plane atomic corrugations are pronounced, leading to an estimated on-site potential variation of several meV. This effect is not observed in RG/hBN with large twist angles ($\theta > 1°$; Fig. 5, E to G), where lattice relaxation is negligible (*47*). Thus, the moiré-modulated spatially sequential filling can be attributed to the prominent non-uniform spontaneous atomic strain in small-twist-angle regime. The electronic structure and Berry curvature distribution of the flat bands are also renormalized by the deformation potential and pseudo-magnetic field arising from this spontaneous atomic strain (*54*). Based on this model, an isolated $|C| = 1$ Chern band can be obtained in small-angle regime (Fig. 5D), offering a platform for realizing the IQAHE and FQAHE. This is absent, however, in the large-angle case (Fig. 5H).

**Discussion and outlook**

We provide direct real-space microscopic visualization of flat-band electronic states in RG moiré superlattices, uncovering a spatial reconstruction behavior governed by moiré periodicity. We attribute the underlying mechanism to atomic-corrugation-induced spatial on-site potential variation at small twist angles, which gives rise to a spatially hierarchical filling within each moiré unit cell. These findings offer several key insights for reexamining correlated and topological phenomena in RG moiré systems. First, electron occupation is spatially non-uniform, exhibiting moiré-modulated orbital textures that depend sensitively on the moiré lattice size. This provides a microscopic foundation for refining theoretical models of correlated and topological phases in these systems. Second, structural relaxation could play an important role in spatial electronic modulation. This can explain the sharp critical threshold of superlattice size (~10 nm) observed here and in transport studies of emergent states such as the QAHE. This again offers crucial constraints for future theoretical efforts. Third, the moiré-modulated hierarchical filling not only renormalizes the electronic structure and topology of the flat bands, but also profoundly reshapes the spatial geometry of correlated states. Since correlation effects are highly sensitive to local filling, this hierarchical mechanism gives rise to moiré-scale spatial modulation of electronic correlations, which can potentially induce diverse correlated-state patterns and thereby facilitate the emergence of exotic quantum phases.

Our findings bear directly on the microscopic understanding of the FQAHE in RG moiré systems. The observation that moiré modulation, and hence the hierarchical filling, emerges above a ~10 nm period—precisely the same threshold above which the FQAHE has been observed in transport—strongly suggests that the spatial renormalization of flat bands is not merely a spectroscopic oddity but a prerequisite for the emergence of topological order. In this picture, the moiré potential serves to create a spatially varying chemical potential landscape that organizes the filling of flat-band electrons in a site-dependent manner. This hierarchical organization, in turn, provides the real-space texture upon which correlation-driven topological phases—including the FQAHE—can nucleate. The fact that the FQAHE appears preferentially on the moiré-distant side of the device is naturally consistent with this picture: it is precisely on this surface that electrons can form an isolated flat band with a nonzero Chern number, as our calculations confirm.

More broadly, our spatially resolved measurements could inspire further exploration of interaction-driven spatial textures in other graphene-based moiré systems with distinct stacking geometries. On the theoretical side, the hierarchical filling mechanism revealed here calls for refined models that explicitly incorporate spatial inhomogeneity and structural relaxation—an essential step toward quantitatively predicting and ultimately engineering the phase diagrams of moiré quantum matter.

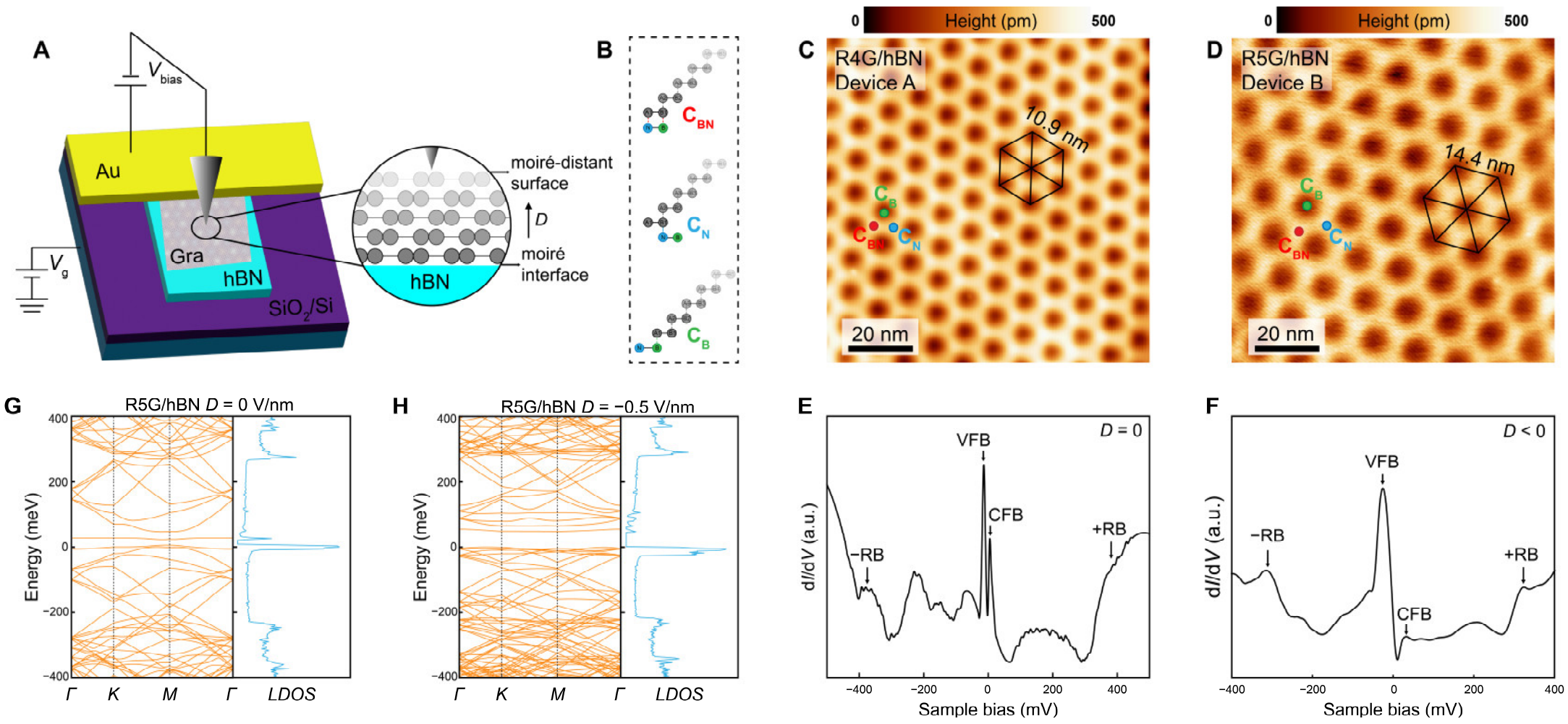

**Fig. 1. Topography and spectroscopy of RG/hBN moiré superlattices.** (**A**) Schematic diagram of STM measurement setup for RG/hBN device. (**B**) Stacking configurations of the RG/hBN moiré superlattice. (**C** and **D**) STM topographic images ($V_b = 0.3$ V, $I = 0.1$ nA) for a R4G/hBN (C) and a R5G/hBN (D) devices. Moiré periods and three high-symmetry stacking sites $C_{BN}$, $C_N$, and $C_B$ are labeled. (**E** and **F**) Typical d$I$/d$V$ spectra of R4G/hBN (E) and R5G/hBN (F) superlattices. +RB (−RB) denotes the conduction (valence) remote band. $V_b = 0.3$ V, $I = 2$ nA, $V_{mod} = 1$ mV, and $V_g = -4$ V (E); $V_b = 0.3$ V, $I = 0.6$ nA, $V_{mod} = 2$ mV, and $V_g = 0$ V (F). (**G** and **H**) Band structures and the corresponding moiré-distant-surface LDOS without (G) and with (H) a displacement field for R5G/hBN superlattices.

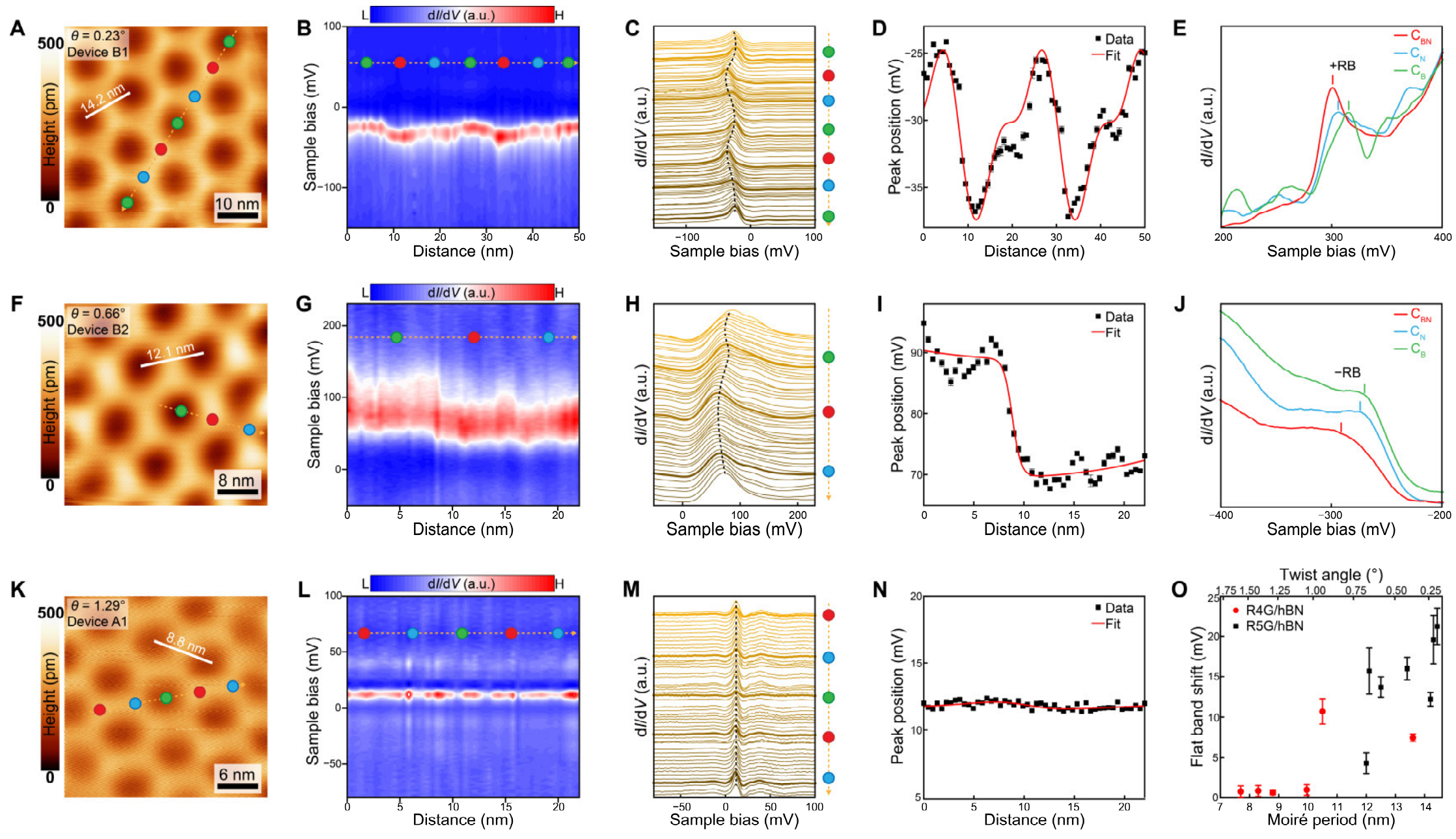

**Fig. 2. Visualization of spatial band renormalization in RG/hBN superlattices.** (**A, F,** and **K**) STM topographic images ($V_b = 0.3$ V, $I = 0.1$ nA) of $\theta = 0.23°$ R5G/hBN (A), $\theta = 0.66°$ R5G/hBN (F), and $\theta = 1.29°$ R4G/hBN (K). (**B**, **G,** and **L**) Contour plots of spatially resolved d$I$/d$V$ spectra measured along the dashed arrows in (A), (F), and (K), respectively. (**C**, **H**, and **M**) Spatially resolved d$I$/d$V$ point spectra corresponding to (B), (G), and (L), respectively. (**D, I**, and **N**) Energy positions of the flat-band peaks extracted from (B), (G), and (L), respectively. The red curves are polynomial fits to the data. (**E** and **J**) Zoom-in high-energy d$I$/d$V$ spectra of three stacking sites for (A) and (F), respectively. +RB/−RB peaks are marked. **O**, Flat-band energy offset as a function of moiré period (twist angle) in R4G/hBN and R5G/hBN superlattices. $V_b = 0.3$ V, $I = 0.6$ nA, $V_{mod} = 2$ mV, and $V_g = 0$ V (B to E); $V_b = 0.3$ V, $I = 0.1$ nA, $V_{mod} = 10$ mV, $V_g = 0$ V, and $T = 77$ K (G to J); $V_b = 0.3$ V, $I = 0.2$ nA, $V_{mod} = 2$ mV, and $V_g = 0$ V (L to N).

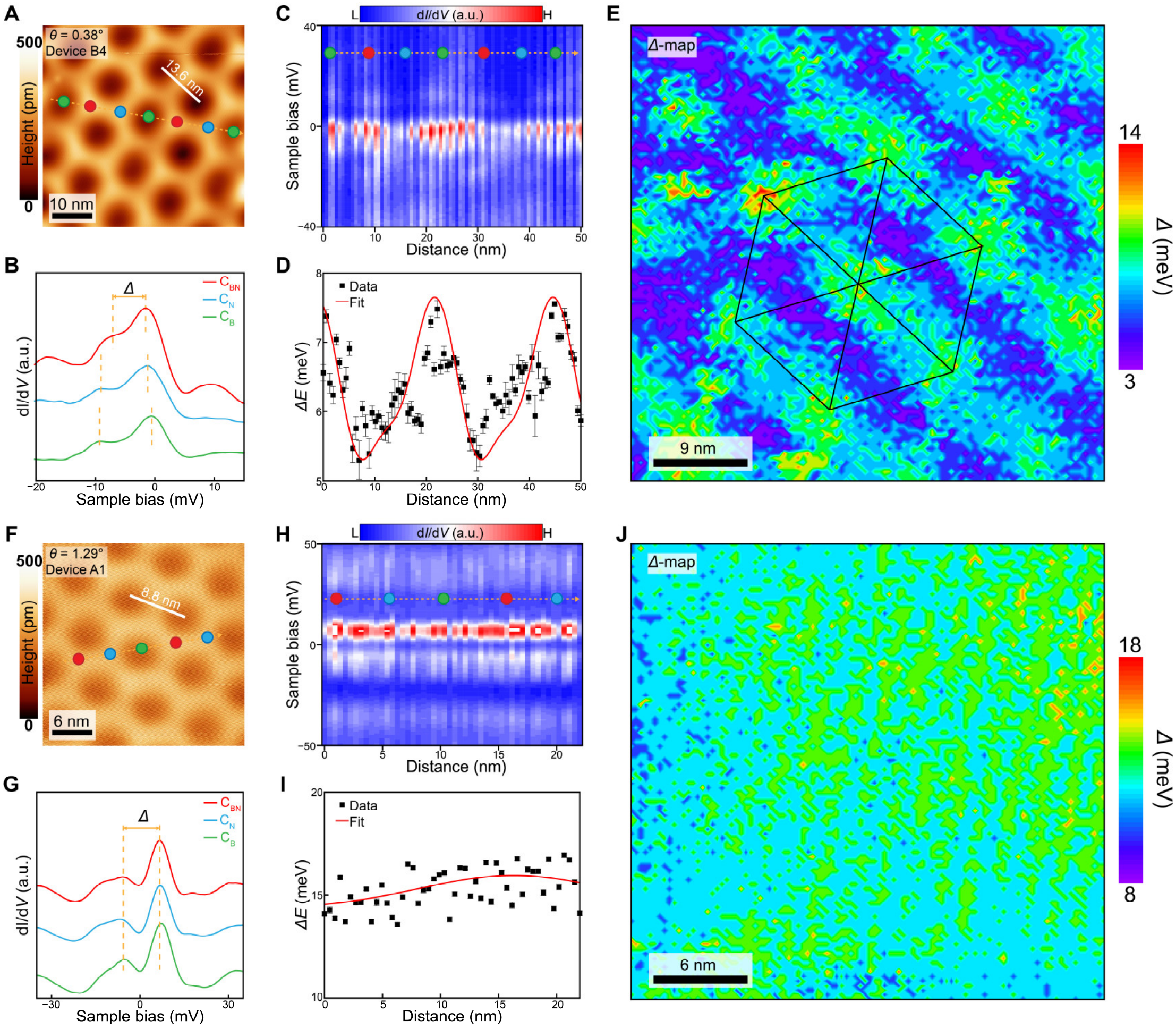

**Fig. 3. Spatial reshaping of electronic correlations.** (**A** and **F**) STM topographic images ($V_b = 0.3$ V, $I = 0.1$ nA) of $\theta = 0.38°$ R5G/hBN (A) and $\theta = 1.29°$ R4G/hBN (F). (**B**) Typical d$I$/d$V$ spectra acquired at three high-symmetry sites in (A), showing partial-filling-induced splitting of the flat band. (**C**) Contour plot of spatially resolved d$I$/d$V$ spectra measured along the arrow in (A). (**D**) Splitting energy of the flat-band peak extracted from (C). (**E**) Spatial distribution diagram ($\Delta$-map) of the splitting values extracted from the sample in (A). (**G** and **H**) d$I$/d$V$ point spectra (G) and spatially resolved contour-plot spectra (H) measured from (F) under partial-filling state of the flat band. (**I**) flat-band splitting energy extracted from (H). The red curves are polynomial fits to the data in (D) and (I). (**J**) $\Delta$-map of flat-band splitting extracted from (F). $V_b = 0.3$ V, $I = 0.6$ nA, $V_{mod} = 2$ mV, and $V_g = 0$ V (B to E); $V_b = 0.3$ V, $I = 0.2$ nA, $V_{mod} = 2$ mV, and $V_g = 21$ V (G to J).

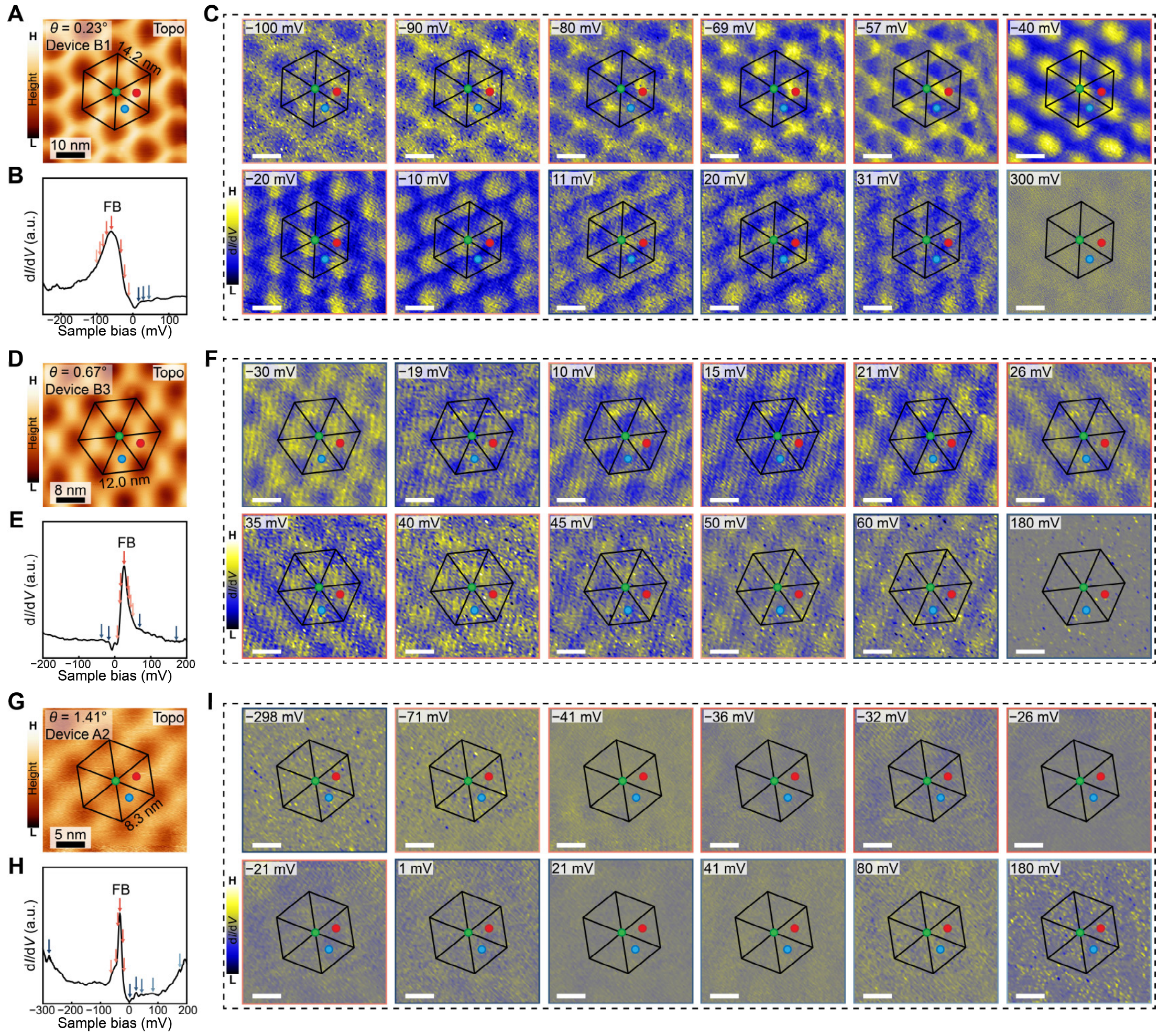

**Fig. 4. Visualization of spatial electronic localization transition**. (**A**, **D**, and **G**) STM topographic images ($V_b = 0.3$ V, $I = 0.1$ nA) of $\theta = 0.23°$ R5G/hBN, $\theta = 0.67°$ R5G/hBN, and $\theta = 1.41°$ R4G/hBN, respectively. (**B**, **E**, and **H**) Typical STS spectra (averaged from three sites) under non-partial fillings for the samples in (A), (D), and (G), respectively. (**C**) d$I$/d$V$ maps acquired at various bias voltages over the same area as in (A). (**F** and **I**) Similar d$I$/d$V$ maps for the samples in (D) and (G), respectively. The measured energy positions of (C), (F), and (I) are marked by corresponding-colored arrows in (B), (E), and (H), respectively. The black hexagons denote the same positions in the images. The red, blue, and green dots indicate the $C_{BN}$, $C_N$, and $C_B$ sites, respectively. $V_b = 0.3$ V, $I = 0.6$ nA, $V_{mod} = 2$ mV, and $V_g = 0$ V.

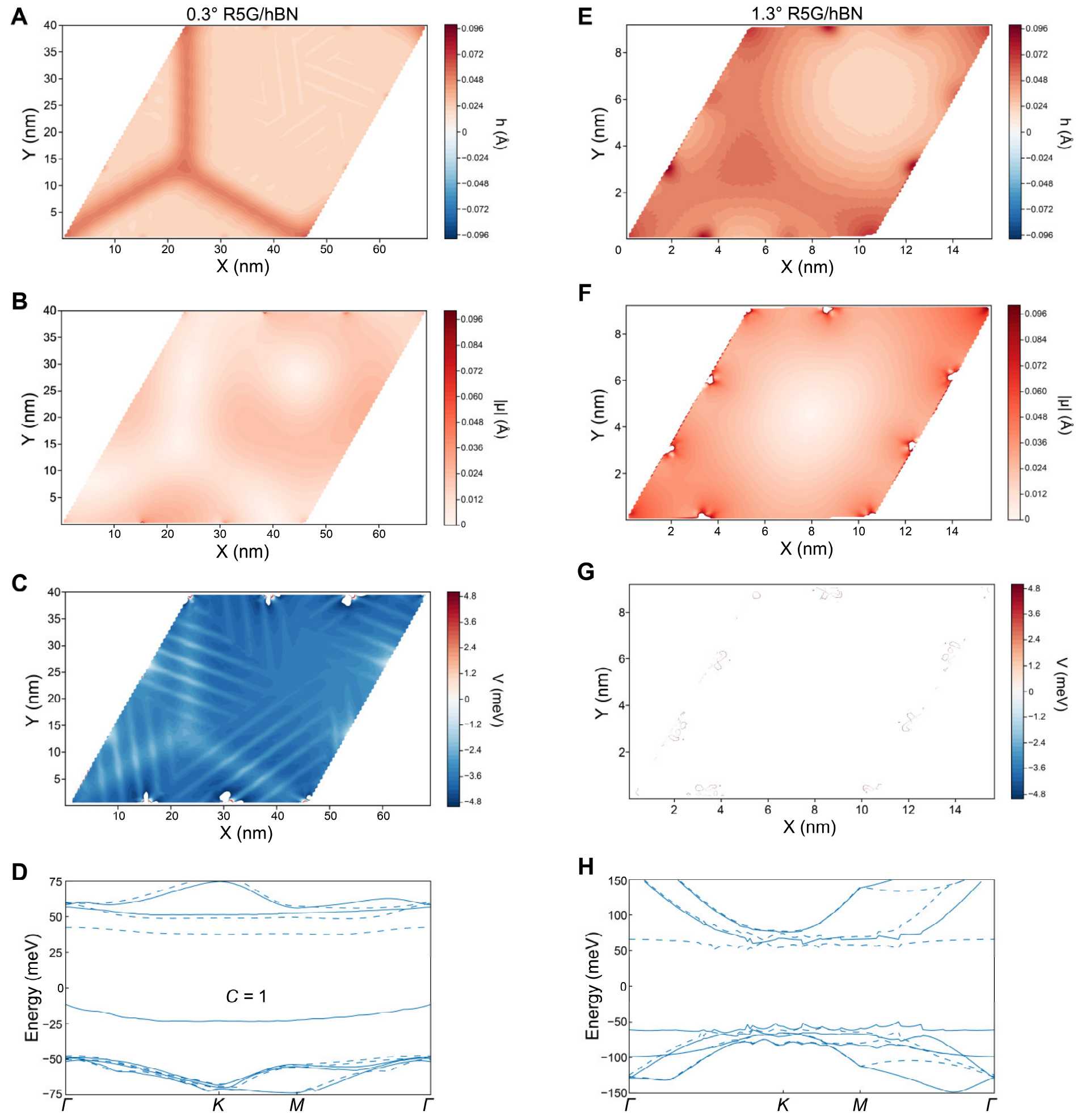

**Fig. 5. Theoretical mechanism of spatial modulation.** (**A** to **C**) Out-of-plane atomic corrugations (A), in-plane atomic corrugations (B) and on-site potential variation (C) in the topmost-layer of a 0.3° R5G/hBN moiré unit cell. (**E** to **G**) Similar calculated results for a 1.3° R5G/hBN moiré unit cell. (**D** and **H**) Hartree-Fock bands with atomic corrugations for 0.3° (D) and 1.3° (H) R5G/hBN at filling $\nu = 1$ and $D = 0.9$ V/nm. Solid/dashed curves represent K/K' valley.

## REFERENCES AND NOTES

1. F. Zhang, B. Sahu, H. Min, A. H. MacDonald, *Phys. Rev. B* **82**, 035409 (2010).
2. N. B. Kopnin, T. T. Heikkilä, G. E. Volovik, *Phys. Rev. B* **83**, 220503(R) (2011).
3. N. B. Kopnin, M. Ijäs, A. Harju, T. T. Heikkilä, *Phys. Rev. B* **87**, 140503(R) (2013).
4. R. Olsen, R. van Gelderen, C. M. Smith, *Phys. Rev. B* **87**, 115414 (2013).
5. D.-H. Xu *et al.*, *Phys. Rev. B* **86**, 201404(R) (2012).
6. B. L. Chittari, G. Chen, Y. Zhang, F. Wang, J. Jung, *Phys. Rev. Lett.* **122**, 016401 (2019).
7. R. Xu *et al.*, *Phys. Rev. B* **91**, 035410 (2015).
8. H. Zhang *et al.*, *Proc. Natl. Acad. Sci. U.S.A.* **121**, e2410714121 (2024).
9. D. Pierucci *et al.*, *ACS Nano* **9**, 5432–5439 (2015).
10. T. Han *et al.*, *Nat. Nanotechnol.* **19**, 181–187 (2023).
11. K. Liu *et al.*, *Nat. Nanotechnol.* **19**, 188–195 (2023).
12. Y. Zhang *et al.*, *Nat. Nanotechnol.* **20**, 222-228 (2025).
13. T. Arp *et al.*, *Nat. Phys.* **20**, 1413–1420 (2024).
14. F. Winterer *et al.*, *Nat. Phys.* **20**, 422-427 (2024).
15. W.-Y. Liao *et al.*, *Phys. Rev. Lett.* **135**, 046202 (2025).
16. A. Kerelsky *et al.*, *Proc. Natl. Acad. Sci. U.S.A.* **118**, e2017366118 (2021).
17. I. Hagymási *et al.*, *Science Advances* **8**, eabo6879 (2022).
18. Y. Liu *et al.*, *Phys. Rev. Lett.* **135**, 156401 (2025).
19. G. Chen *et al.*, *Nature* **572**, 215–219 (2019).
20. H. Zhou, T. Xie, T. Taniguchi, K. Watanabe, A. F. Young, *Nature* **598**, 434-438 (2021).
21. C. L. Patterson *et al.*, *Nature* **641**, 632-638 (2025).
22. T. Han *et al.*, *Nature* **643**, 654-661 (2025).
23. J. Yang *et al.*, *Nat. Mater.* **24**, 1058–1065 (2025).
24. G. Chen *et al.*, *Nature* **579**, 56-61 (2020).
25. H. Zhou *et al.*, *Nature* **598**, 429-433 (2021).
26. W. Zhou *et al.*, *Nat. Commun.* **15**, 2597 (2024).
27. T. Han *et al.*, *Nature* **623**, 41-47 (2023).
28. Y. Lee *et al.*, *Nano Lett.* **22**, 5094-5099 (2022).
29. Z. Lu *et al.*, *Nature* **626**, 759-764 (2024).
30. Y. Sha *et al.*, *Science* **384**, 414-419 (2024).
31. T. Han *et al.*, *Science* **384**, 647-651 (2024).
32. Y. Choi *et al.*, *Nature* **639**, 342-347 (2025).
33. Z. Lu *et al.*, *Nature* **637**, 1090-1095 (2025).
34. J. Xie *et al.*, *Nat. Mater.* **24**, 1042–1048 (2025).
35. J. Cai *et al.*, *Nature* **622**, 63-68 (2023).
36. H. Park *et al.*, *Nature* **622**, 74-79 (2023).
37. Y. Zeng *et al.*, *Nature* **622**, 69-73 (2023).
38. F. Xu *et al.*, *Phys. Rev. X* **13**, 031037 (2023).
39. Z. Dong, A. S. Patri, T. Senthil, *Phys. Rev. Lett.* **133**, 206502 (2024).
40. B. Zhou, H. Yang, Y.-H. Zhang, *Phys. Rev. Lett.* **133**, 206504 (2024).
41. J. Dong *et al.*, *Phys. Rev. Lett.* **133**, 206503 (2024).
42. Y. H. Kwan *et al.*, *Phys. Rev. B* **112**, 075109 (2025).
43. X. Lu, Y. Yang, Z. Guo, J. Liu, *arXiv: 2509.19764* (2025).

44. J. Jung, A. Raoux, Z. Qiao, A. H. MacDonald, *Phys. Rev. B* **89**, 205414 (2014).
45. G. Argentero *et al.*, *Nano Lett.* **17**, 1409-1416 (2017).
46. S. Zhou, J. Han, S. Dai, J. Sun, D. J. Srolovitz, *Phys. Rev. B* **92**, 155438 (2015).
47. C. R. Woods *et al.*, *Nat. Phys.* **10**, 451-456 (2014).
48. J. Jung, A. M. DaSilva, A. H. MacDonald, S. Adam, *Nat. Commun.* **6**, 6308 (2015).
49. G. M. Rutter *et al.*, *Nat. Phys.* **7**, 649-655 (2011).
50. M. Yankowitz, F. Wang, C. N. Lau, B. J. LeRoy, *Phys. Rev. B* **87**, 165102 (2013).
51. L.-J. Yin *et al.*, *Phys. Rev. B* **95**, 081402(R) (2017).
52. L.-J. Yin *et al.*, *Phys. Rev. Lett.* **122**, 146802 (2019).
53. E. Seewald *et al.*, *arXiv:2510.09548* (2025).
54. L. Nashabeh, H. Ochoa, *arXiv:12605.16218* (2026).

**ACKNOWLEDGMENTS**

We thank J.Y. Hu and J. Dai for assistance with a piece of related equipment for low-temperature measurement. **Funding:** This work was supported by the National Natural Science Foundation of China (Grant Nos. 12474166, 12174095, 12474167, 12174096, 12304217, 12204164, and 12574188) and the Natural Science Foundation of Hunan Province, China (Grant Nos. 2025JJ20001, 2025JJ60002, and 2025JJ50020). L.-J.Y. acknowledges support from the Scientific Research Innovation Capability Support Project for Young Faculty, China (Grant No. SRICSPYF-ZY2025086) and Project of Yuelushan Center for Industrial Innovation (Grant No. 2025YCII0208). We also acknowledge the financial support from the Fundamental Research Funds for the Central Universities of China.

**Author contributions:** P.-C.P. and Y.Z. fabricated the samples with the help of L.-H.T., Y.T., and Y.H. P.-C.P. performed the STM experiments with the help of Y.Z., J.H., and C.-C.X. Li Z. and Lijie Z. contributed to the experimental discussions. P.-C.P. and L.-J.Y. analyzed the data. S.Z. performed the theoretical calculations. W.-X.W., Z.Q., and L.-J.Y. supervised the experiments. L.-J.Y. conceived the project. P.-C.P., S.Z., and L.-J.Y. wrote the manuscript with input from all other authors.

**Competing interests:** The authors declare no competing interests.

**Data and materials availability:** All data are available in the manuscript or the supplementary materials. All data needed to evaluate the conclusions in this paper are present in the main text or the supplementary materials.

**SUPPLEMENTARY MATERIALS**

Materials and Methods

Supplementary Text

Figs. S1 to S15

References (55-64)